\documentstyle[12pt]{article}

\newlength{\titlesep}
\newlength{\authorsep}
\makeatletter

\newcount\@arga
\newcount\@argb
\newcount\@argc
\newcount\@ctmpa
\newcount\@ctmpb
\newcount\@ctmpc
\newcount\@ctmpd
\newcount\@ctmpe
\newdimen\@darg
\newdimen\@bblen
\newif\if@bbllx
\newif\if@bblly
\newif\if@bburx
\newif\if@bbury
\newif\if@height
\newif\if@width
\newif\if@scale
\newif\ifno@bb
\newif\ifepsfdraft
\epsfdraftfalse
\def\@setpsfile#1{
                \typeout{epsf:[#1]}
                \def\@psfile{#1}
}
\def\@setpsheight#1{
                \@heighttrue
                \@darg=#1\relax
                \edef\@psheight{\number\@darg}
}
\def\@setpswidth#1{
                \@widthtrue
                \@darg=#1\relax
                \edef\@pswidth{\number\@darg}
}
\def\@setpsscale#1{
                \@scaletrue
                \def\@pshscale{#1}
                \def\@psvscale{#1}
                \@bblen=#1pt\relax
                \@bblen=1000\@bblen
                \edef\@texhscale{\expandafter\remove@dim\the\@bblen}
                \let\@texvscale=\@texhscale
}
\def\@setpshscale#1{
                \@scaletrue
                \def\@pshscale{#1}
                \@bblen=#1pt\relax
                \@bblen=1000\@bblen
                \edef\@texhscale{\expandafter\remove@dim\the\@bblen}
}

\def\@setpsvscale#1{
                \@scaletrue
                \def\@psvscale{#1}
                \@bblen=#1pt\relax
                \@bblen=1000\@bblen
                \edef\@texvscale{\expandafter\remove@dim\the\@bblen}
}

\def\@setparms#1=#2,{\@nameuse{@setps#1}{#2}}
\def\ps@init@parms{
                \@heightfalse \@widthfalse
                \no@bbfalse
                \def\@psbbllx{}\def\@psbblly{}
                \def\@psbburx{}\def\@psbbury{}
                \def\@psheight{}\def\@pswidth{}
                \def\@pshscale{1}\def\@psvscale{1}
                \def\@texhscale{1000}\def\@texvscale{1000}
                \def\@psfile{}
                \def\@sc{}
}
\def\parse@ps@parms#1{
                \@for\@epsfile:=#1\do
                   {\expandafter\@setparms\@epsfile,}}
\newif\ifnot@eof
\newread\ps@stream
\def\bb@search{
        \openin\ps@stream=\@psfile
        \no@bbtrue
        \not@eoftrue
        \catcode`\%=12\relax
        \ifeof\ps@stream\typeout{epsf: File not found}\fi
        \loop
                \read\ps@stream to \line@in
                \global\toks200=\expandafter{\line@in}\relax
                \ifeof\ps@stream \not@eoffalse \fi
                \@bbtest{\toks200}\relax
                \if@bbmatch\not@eoffalse\expandafter\bb@cull\the\toks200\fi
        \ifnot@eof \repeat
        \catcode`\%=14
}       

\catcode`\%=12
\newif\if@bbmatch
\def\@bbtest#1{\expandafter\@a@\the#1%%BoundingBox:\@bbtest\@a@}
\long\def\@a@#1%%BoundingBox:#2#3\@a@{
        \ifx\@bbtest#2\@bbmatchfalse\else\@bbmatchtrue\fi}
\def\bb@cull %%BoundingBox:{
        \@ifnextchar\space{\@latexbug}{\bb@extract}}
\def\bb@extract #1 #2 #3 #4 {
        \message{BoundingBox: (#1bp,#2bp)--(#3bp,#4bp)}
        \@darg=#1 bp\edef\@psbbllx{\number\@darg}
        \@darg=#2 bp\edef\@psbblly{\number\@darg}
        \@darg=#3 bp\edef\@psbburx{\number\@darg}
        \@darg=#4 bp\edef\@psbbury{\number\@darg}
        \no@bbfalse
}
\catcode`\%=14

\def\compute@bb{
                \bb@search
                \ifno@bb \typeout{epsf: No BoundingBox}
                \stop
                \else
                \@arga=\@psbburx
                \advance\@arga by -\@psbbllx
                \edef\@bbw{\number\@arga}
                \@arga=\@psbbury
                \advance\@arga by -\@psbblly
                \edef\@bbh{\number\@arga}
                \fi
}
\def\in@hundreds#1#2#3{\@argb=#2 \@argc=#3
                     \@ctmpa=\@argb     % @ctmpa is first digit #2/#3
                     \divide\@ctmpa by \@argc
                     \@ctmpb=\@ctmpa
                     \multiply\@ctmpb by \@argc
                     \advance\@argb by -\@ctmpb
                     \multiply\@argb by 10
                     \@ctmpb=\@argb     % @ctmpb is second digit of #2/#3
                     \divide\@ctmpb by \@argc
                     \@ctmpc=\@ctmpb
                     \multiply\@ctmpc by \@argc
                     \advance\@argb by -\@ctmpc
                     \multiply\@argb by 10
                     \@ctmpc=\@argb     % @ctmpc is the third digit
                     \divide\@ctmpc by \@argc
                     \@arga=#1\@ctmpe=0
                     \@ctmpd=\@arga
                        \multiply\@ctmpd by \@ctmpa
                        \advance\@ctmpe by \@ctmpd
                     \@ctmpd=\@arga
                        \divide\@ctmpd by 10
                        \multiply\@ctmpd by \@ctmpb
                        \advance\@ctmpe by \@ctmpd
                     \@ctmpd=\@arga
                        \divide\@ctmpd by 100
                        \multiply\@ctmpd by \@ctmpc
                        \advance\@ctmpe by \@ctmpd
                     \edef\@result{\number\@ctmpe}
}
\def\compute@wfromh{
                \in@hundreds{\@psheight}{\@bbw}{\@bbh}
                \edef\@pswidth{\@result}
}
\def\compute@hfromw{
                \in@hundreds{\@pswidth}{\@bbh}{\@bbw}
                \edef\@psheight{\@result}
}
\def\compute@handw{
        \if@height 
                \if@width
                \else
                        \compute@wfromh
                \fi
        \else 
                \if@width
                        \compute@hfromw
                \else
                        \if@scale
                                \in@hundreds{\@texvscale}{\@bbh}{1000}
                                \let\@bbh=\@result
                                \in@hundreds{\@texhscale}{\@bbw}{1000}
                                \let\@bbw=\@result
                        \fi
                                \edef\@psheight{\@bbh}
                                \edef\@pswidth{\@bbw}
                \fi
        \fi
}
{\catcode`\p=12\catcode`\t=12
\gdef\remove@dim#1.#2pt{#1}}

\def\compute@sizes{
        \compute@bb
        \compute@handw
}
\def\epsfile#1{
        \ps@init@parms
        \parse@ps@parms{#1}
        \compute@sizes
        \@arga=\@psheight
        \divide\@arga by 65536
        \edef\@psvsize{\number\@arga}
        \@arga=\@pswidth
        \divide\@arga by 65536
        \edef\@pshsize{\number\@arga}
        \message{=>(\@pshsize bp,\@psvsize bp)}
        \leavevmode
        \vbox to \@psheight true sp{
                \hbox to \@pswidth true sp{
                \ifepsfdraft\hss\@psfile\hss\else
                \if@height 
                        \if@width
                                \special{epsfile=\@psfile \space 
                                hsize=\@pshsize \space
                                vsize=\@psvsize \space}
                        \else
                                \special{epsfile=\@psfile \space 
                                vsize=\@psvsize \space}
                        \fi
                \else 
                        \if@width
                                \special{epsfile=\@psfile \space 
                                hsize=\@pshsize \space}
                        \else
                                \if@scale
                                        \special{epsfile=\@psfile \space
                                        vscale=\@psvscale \space
                                        hscale=\@pshscale \space}
                                \else
                                        \special{epsfile=\@psfile \space}
                                \fi
                        \fi
                \fi
                \hfil\fi
                }
        \vfil
        }
}

\renewcommand{\thesection}{\Roman{section}}
\@addtoreset{equation}{section}
\renewcommand{\appendix}{\par
  \setcounter{section}{0}
  \setcounter{subsection}{0}
  \renewcommand{\thesection}{Appendix~\Alph{section}}
  \renewcommand{\theequation}{\Alph{section}.\arabic{equation}}}

\def\fnum@figure{FIG.~\thefigure}
\newcommand{\Fig}[1]{\item[{\bf FIG.~\protect\ref{#1}}]}

\newcounter{figureparent}
\@addtoreset{figure}{figureparent}

\newcounter{eqnparent}
\@addtoreset{equation}{eqnparent}

\renewcommand{\abstract}{\if@twocolumn
  \section*{Abstract}
  \else
  \begin{center}
    {\bf Abstract\vspace{-.5em}\vspace{0pt}} 
  \end{center}
  \quotation 
  \fi}
\renewcommand{\endabstract}{\if@twocolumn\else\endquotation\fi}

\newcommand{\thismonth}{\ifcase\month\or
 January\or February\or March\or April\or May\or June\or
 July\or August\or September\or October\or November\or December\fi
 \space \number\year}

\newcommand{\preprintnumber}[1]
{\begin{flushright}
  \begin{tabular}{l} #1 \end{tabular}
  \end{flushright}}

\makeatother

\newcommand{\rn}[1]{{\romannumeral#1}}
\newcommand{\Rn}[1]{{\uppercase\expandafter{\romannumeral#1}}}
\newcommand{\gsim}%
{\mbox{\raisebox{-1.0ex}
    {$\ \stackrel{\textstyle >}{\textstyle \sim}\ $}}}
\newcommand{\lsim}%
 {\mbox{\raisebox{-1.0ex}
     {$\ \stackrel{\textstyle <}{\textstyle \sim}\ $}}}
\newcommand{\ie}{{\it i.e.}\ }
\newcommand{\etal}{{\it et al.\/}}
\newcommand{\vev}[1]{\left\langle #1 \right\rangle}
\newcommand{\ol}[1]{\overline{#1}}
\newcommand{\bsg}{$b\rightarrow s\ \gamma$\ }
\newcommand{\mbsg}{\ b\rightarrow s\ \gamma\ }
\newcommand{\br}{{\rm Br}}

\newcommand{\Journal}[4]{{#1} {\bf #2} {(#3)} {#4}}

\newcommand{\plb}{\sl Phys.~Lett.~{\bf B}}
\newcommand{\prp}{\sl Phys.~Rep.}

\newcommand{\prd}{\sl Phys.~Rev.~{\bf D}}
\newcommand{\prl}{\sl Phys.~Rev.~Lett.}

\newcommand{\npb}{\sl Nucl.~Phys.~{\bf B}}

\newcommand{\zpc}{\sl Z.~Phys.~{\bf C}}

\begin{document}
\baselineskip 18pt

\begin{titlepage}
\preprintnumber{%
KEK-TH-421\\
KEK preprint 94-159\\
HUPD-9417\\
December 1994
}
\vspace*{\titlesep}
\begin{center}
{\LARGE\bf Charged Higgs mass bound from the \bsg process in
  the minimal supergravity model}\\
\vspace*{\titlesep}
{\large Toru {\sc Goto}$^1$} and {\large Yasuhiro {\sc Okada}$^2$}\\
\vspace*{\authorsep}
{\it $^1$Department of Physics, Hiroshima University \\
  1-3-1 Kagamiyama, Higashi Hiroshima, 724 Japan}\\
\vspace*{\authorsep}
{\it $^2$Theory Group, KEK, Tsukuba, Ibaraki, 305 Japan }
\end{center}
\vspace*{\titlesep}
\begin{abstract}
We study the constraint on the 
mass of the charged Higgs boson in the minimal supergravity model 
based on the recent measurement of the inclusive \bsg decay. 
It is shown that the lower bound for the charged Higgs mass
crucially depends on the sign of the higgsino
mass parameter ($\mu$).
For $\mu<0$, the bound exceeds 400 GeV
when the ratio of two Higgs vacuum expectation values
($\tan\beta$) is larger than 10.  
No strong bound is obtained for $\mu>0$ due to
cancellations between charged Higgs and supersymmetric 
particle contributions.
For $3\lsim\tan\beta\lsim5$, a charged Higgs lighter 
than 180 GeV is excluded
by this process irrespective of the sign of $\mu$.
\end{abstract}

\end{titlepage}

%%%%%%%%%%%%%%%%%%%%%%%%%%%%%%%%%%%%%%%%%%%%%%%%%%%%%%%%%%%%%%%%%%%%%%

Flavor changing neutral current (FCNC) processes play a unique role 
in searching for physics beyond the standard model (SM) of 
elementary particles.  
These processes are sensitive to virtual effects of new particles, 
since the FCNC processes in SM do not occur at the tree level. 
These processes can thus be more powerful than direct particle searches 
in putting constraints on the parameter space of various new physics. 
In particular, the radiative decay of the $b$ quark, \bsg, deserves 
a special attention.  
It has been noticed that in a two Higgs doublet model (THDM) the
charged Higgs boson can give a substantial contribution to
the \bsg rate \cite{thdm,GSW}.
Recently, the CLEO group \cite{cleo} has reported the first
measurement of the inclusive \bsg branching ratio
$  \br( \mbsg ) = ( 2.32 \pm 0.51 \pm 0.29 \pm 0.32)
                 \times 10^{-4} ~$.  
In fact, this result constrains the mass of the charged Higgs
boson in a certain type of THDM called Model \Rn{2} \cite{thdm,GSW}
to be larger than 260 GeV \cite{cleo}.

The Higgs sector in the minimal supersymmetric (SUSY) extension of SM 
is a special case of Model \Rn{2} THDM.
However, the above-mentioned limit cannot be directly applied, because 
SUSY particles can contribute to the \bsg process in 
addition to the SM particles and the charged Higgs.
It is a natural question to ask how the charged Higgs mass limit is modified 
in the SUSY extension of SM.  
Many authors have discussed the \bsg process in SUSY \cite{BBMR}-\cite{Carena}.
Although the SM and the charged Higgs contribution to the 
\bsg amplitude has the same sign, it is found that the SUSY loop 
can interfere either constructively or destructively with them. 
The limit for the charged Higgs mass from 
this process can be weakened by the effects of the SUSY particles.

The minimal supergravity model provides an attractive framework for 
SUSY extension of SM.
In this model, masses and mixing parameters for SUSY
particles can be expressed by a few soft SUSY breaking
parameters as well as gauge and Yukawa coupling constants.
The \bsg branching ratio thus depends on much smaller
number of free parameters compared to that in general
SUSY standard models.
It has been noticed \cite{Lopez1,BertoliniVissani,Lopez2} 
that the sign of the SUSY loop contributions with respect to
those of the SM and charged Higgs is strongly correlated with
the sign of the higgsino mass parameter \ie the $\mu$
parameter in the minimal supergravity model.

In this paper, we compare the CLEO data with the prediction
of the minimal supergravity model and determine the allowed
region in the parameter space of the model.
Scanning the free parameter space extensively, we search for the
constraint on the charged Higgs mass.
It will be shown that the lower bound of the charged Higgs
mass crucially depends on the sign of $\mu$.
The bound becomes much larger than that in the non-SUSY
Model \Rn{2} THDM for $\mu<0$, while no strong bound is
obtained for $\mu>0$.

The calculation of the \bsg branching ratio has already been 
discussed extensively in the literature
\cite{GSW,Misiak,BurasMisiakMunzPokorski}.
The decay rate for \bsg normalized to the semileptonic decay rate 
is given by
\begin{eqnarray}
  \frac{\Gamma(\mbsg)}{\Gamma(\ b \rightarrow c\ e\ \ol{\nu}\ )}
  &=& \frac{1}{|V_{cb}|^2}
      \frac{6\alpha_{\rm QED}}{\pi g(m_c/m_b)}
      \left| V_{ts}^* V_{tb} C_7^{\rm eff}(Q) \right|^2 ~,
  \label{BRformula}
\\
    C_7^{\rm eff}(Q)
    &=& \eta^{16/23} C_7(M_W)
       + \frac{8}{3}\left(   \eta^{14/23}
                           - \eta^{16/23} \right) C_8(M_W)
       + C ~,
\nonumber
\end{eqnarray}
where $\eta = \alpha_s(M_W)/\alpha_s(Q)$, $Q$ being the scale 
of the order of the bottom mass, 
and $g(z) = 1 - 8z^2 + 8z^6 - z^8 - 24z^4\ln z$.  
$C$ is a constant which depends on $\eta$.
The above formula takes the leading order QCD corrections into account.
The $C_7(M_W)$ and $C_8(M_W)$ are coefficients of the
magnetic and chromomagnetic operators at $M_W$.
The $C$ term is induced by operator mixing in evolving 
from $M_W$ to the low energy scale $Q$.

Ambiguities in the calculation are discussed in detail in Ref.\
\cite{BurasMisiakMunzPokorski}.
The most important ambiguity comes from the choice of the
renormalization scale $Q$.
Varying $Q$ between $m_b/2$ and $2 m_b$ induces an ambiguity
of $\pm 25\ \%$ for the branching ratio in SM.
Other ambiguities include the choice of $m_c/m_b$ (which affects 
the semileptonic rate) and the value of $\alpha_s(M_Z)$.

The coefficients $C_7(M_W)$ and $C_8(M_W)$ receive the following 
contributions at one loop:
(\rn{1}) the $W$ and top quark loop,
(\rn{2}) the charged Higgs and top quark loop,
(\rn{3}) the chargino and up-type squark loops,
(\rn{4}) the gluino and down-type squark loops and 
(\rn{5}) the neutralino and down-type squark loops.
The contribution from (\rn{5}) is known to be very 
small \cite{BBMR}, which we will ignore hereafter.
The charged Higgs contribution depends on its mass and the
ratio of the vacuum expectation values of the two Higgs doublets,
\ie $\tan\beta =
\vev{H_2^0} / \vev{H_1^0}$, where $H_1^0$ and $H_2^0$ are
the neutral component of the two Higgs doublets.
The chargino and gluino loop contributions depend on the mass
and mixing of the particles inside the loop.
Although the squark mixing matrices are arbitrary parameters 
in a general SUSY standard model, 
they can be calculated from the flavor mixing matrix of quarks 
(the Cabibbo-Kobayashi-Maskawa matrix) in the minimal supergravity model
by solving the renormalization group equations for
various soft SUSY breaking terms.

The soft SUSY breaking parameters at the GUT
scale are the universal scalar mass ($m_0$), a parameter in
the trilinear coupling of scalars ($A_X$), a parameter in
the two Higgs coupling ($B_X$) and the gaugino mass ($M_{gX}$).
We are assuming the GUT relation for the three gaugino masses \ie the
SU(3), SU(2) and U(1) gaugino mass parameters are equal at the GUT scale.
Besides these soft SUSY breaking parameters, the
superpotential contains the Yukawa coupling constants and
the $\mu$ parameter. 
Given a value for the top mass and $\tan\beta$, we determine all the
particle masses and mixings at the weak scale by solving
relevant renormalization group equations with initial
conditions at the GUT scale specified by the above parameters.
We compute the Higgs effective potential at the weak scale 
and require that the electroweak symmetry is
broken properly (the radiative breaking scenario).
We include the one loop corrections to the effective potential 
induced by the Yukawa couplings for the third generation. 
The condition for radiative breaking with the correct scale 
reduces the number of free
parameters to three for given $\tan\beta$ and $m_t$. 
We can think of these parameters as the charged Higgs mass
($m_{H^\pm}$), $\mu$ and $M_2$ (SU(2) gaugino mass) at the weak scale.
For a given set of these five parameters, 
all other masses and mixings are thus calculable.  

We now present the results of the \bsg branching ratio.
Besides the radiative breaking condition we require the following
phenomenological constraints \cite{pdg}:
\begin{enumerate}
\item The mass of any charged SUSY particle is larger than 45 GeV;
\item The sneutrino mass is larger than 41 GeV;
\item The gluino mass is larger than 100 GeV;
\item Neutralino search results at LEP \cite{aleph}, which require
$\Gamma(Z \rightarrow \chi \chi)< 22$ MeV,
$\Gamma(Z \rightarrow \chi \chi')$,
$\Gamma(Z \rightarrow \chi' \chi')< 5 \times 10^{-5}$ GeV,
where $\chi$ is the lightest neutralino and 
$\chi'$ is any neutralino other than the lightest one;
\item The lightest SUSY particle (LSP) is neutral;
\item The condition for not having a charge or color
  symmetry breaking vacuum \cite{aterm}.
\end{enumerate}
In Fig.\ \ref{fig:br}, we show the \bsg branching ratio for
$m_t = 175$ GeV%
\footnote{%
This top quark mass is the $\overline{\rm MS}$ running mass
at $Q=M_Z$.
This mass coincides with the pole mass within a few percent
\cite{BurasJaminWeisz}.
}
and $\tan\beta = 5$.
Each point in the figure corresponds to the value of the 
\bsg branching ratio for each scanned point in the parameter
space compatible with the above conditions.
This branching ratio includes the chargino and gluino loop
contributions as well as the SM and the charged Higgs loop. 
The line in the figure represents the branching ratio when 
only the SM and charged Higgs contributions are retained.
We notice that the points are divided by this line.
In fact, the points above and below this line correspond
to $\mu < 0$ and $\mu > 0$ cases respectively%
\footnote{
Our convention of the sign of $\mu$ is opposite to those 
in Refs.\ \cite{BertoliniVissani} and \cite{Lopez2}.
}.
This confirms the assertion \cite{Lopez1,BertoliniVissani,Lopez2} that 
the sign of $\mu$ determines whether the SUSY contribution enhances or
suppresses the \bsg branching ratio.

We show the excluded region in the $\tan\beta$ and $m_{H^\pm}$ space 
in Fig.~\ref{fig:minus} and Fig.\ \ref{fig:plus}.
The two branches $\mu>0$ and $\mu<0$ are separately plotted. 
The  excluded region is determined using the CLEO
result  $1 \times 10^{-4} < \br(\mbsg) < 4 \times 10^{-4}$.
In order to take account of the theoretical uncertainties
we have calculated the \bsg branching ratio by varying the
renormalization scale $Q$ between $m_b/2$ and $2 m_b$. 
To be conservative, we also allow an additional 10\% uncertainty.
In the calculation we have used $ \alpha_s(m_Z)=0.116$,
$m_c/m_b=0.3$ and $m_b=4.25$ GeV. 
We regard a point in $(\ \tan\beta,\ m_{H^\pm}\ )$ space
excluded when the branching ratio cannot be within the CLEO
bound for any choice of other parameters (\ie $\mu$ and
$M_2$) even if we consider the above-mentioned theoretical
ambiguities.
We also show in these figures the lower bound of the charged
Higgs mass when only the SM and the charged Higgs contributions
are retained.  In comparison, 
the excluded region in minimal supergravity becomes larger when $\mu
< 0$. The bound reaches 400 GeV for $\tan \beta > 10$.
For $\mu > 0$, the \bsg process is not very effective in 
constraining the charged Higgs mass,
because the charged Higgs contribution can be completely
cancelled by SUSY particle contributions.

We combine the two branches in Fig.~\ref{fig:exclude}.  
The region where $3 \lsim\tan\beta \lsim 5$ and the charged
Higgs is lighter than 180 GeV is excluded 
irrespective of the sign of $\mu$%
\footnote{
A similar observation was made in Ref.\
\cite{BertoliniVissani} for a special choice of the $B$ parameter.
}.
Before including the \bsg constraint, 
this region was allowed only for $\mu < 0$ because of the
phenomenological and radiative breaking conditions.  
The \bsg process now completely excludes this region.

The reason for the strong dependence on $\mu$ can be understood
as follows. For the chargino and up-type squark loop the most
important contribution comes from the loop diagram with the top and
bottom Yukawa coupling constants. This diagram is proportional
to a product of the left-right mixing parameter of the
stops,\ie $A_t-\mu \cot \beta$, and the higgsino mixing parameter
$\mu$. The parameter $A_t$ is determined by $A_X$ 
and $M_{gX}$ through the renormalization
group equations. An interesting observation is that for such
a high value of the top quark mass as considered here $A_t$ 
is almost independent of $A_X$ and proportional to $M_{gX}$.
Moreover, the $\mu \cot \beta $ term is suppressed for  $\tan\beta 
> 3$.  Therefore, the amplitude from this sector is proportional
to a product of the gaugino mass and the higgsino mass parameter
in a very good approximation. A similar
consideration applies to the gluino and down-type squark loop.
In this case a sizable contribution can arise when the graph
involves the left-right mixing in the sbottom sector, especially
for a large value of $\tan\beta$. Then, the amplitude is proportional
to a product of the gluino mass $(M_3)$ and the sbottom mixing
parameter, \ie  $A_b-\mu \tan \beta$. For a large value of $\tan\beta$
the latter parameter is governed by the $\mu \tan \beta$ term.
Here again, the contribution to the amplitude is almost 
proportional to the product of the gaugino mass and the higgsino 
mass parameter. In both cases, the SUSY
contribution enhances (suppresses) the SM amplitude when
$\mu < 0$ ($\mu > 0$).
 
Let us discuss generalizations of our results.
First, we consider the situation when LEP \Rn{2} fails to
find any SUSY particle.
Then the lower bounds for the charged SUSY particle masses rise
to about 90 GeV.
The excluded region of the charged Higgs mass and
$\tan\beta$ are shown in Fig.\ \ref{fig:lepIIminus} and Fig.\
\ref{fig:lepIIplus}.
We can see that the allowed region by the phenomenological
constraint and radiative breaking condition is shifted to
higher values of the charged Higgs mass.
For $\mu<0$, the line for the lower bound from the \bsg
constraint is unchanged and still a large portion of the
parameter space is excluded.
Therefore the present \bsg constraint can be stronger than the
constraint provided by the LEP \Rn{2} search in determining
an allowed parameter region in the minimal
supergravity model.

Next, we would like to study how strongly the results depend on
the assumption of the universal scalar mass at the GUT scale.
By defining the tree Higgs potential as;
\begin{equation}
  V_{\rm tree}=(\Delta_1^2+\mu^2)|H_1|^2+(\Delta_2^2+\mu^2)|H_2|^2 
            -B\mu(H_1 H_2 + H_1^* H_2^*) + (D~ {\rm terms}),
\end{equation}
we have changed the initial condition for the Higgs soft
breaking term, $\Delta_1^2$, $\Delta_2^2$ as
$\Delta_2^2=r\cdot m_0^2$, $\Delta_1^2=m_0^2$, where $m_0^2$
is an universal scalar mass for squarks and sleptons. We
have introduced a parameter $r$ to relax the assumption that
all the scalar fields have a common soft SUSY breaking mass.
In this way we can change the parameter space allowed by the
requirement of the radiative electroweak symmetry breaking.
Although the allowed parameter region in $\mu$-$M_2$ space
is changed for various choice of $r$ (we varied $r$ between
0.1 and 10),
the situation that
the charged Higgs mass is strongly constrained only for
$\mu<0$ remains true for any choice of $r$.
As an example, the bound on the charged Higgs mass from the
\bsg process is shown in Fig.\ \ref{fig:rminus} and Fig.\
\ref{fig:rplus} for the choice $r=4$.
These analyses suggest that our results are independent of
the details of the radiative breaking mechanism.

Finally, we will consider how the results depends on the top
mass.
It turns out that the correlation between the sign of $\mu$
and the sign of the \bsg amplitude remains unchanged,
although the numerical value of the charged Higgs mass bound
depends on $m_t$.
In the case of $m_t = 150$ GeV, where the lower bound of the
charged Higgs mass without SUSY contributions is about 200
GeV, the charged Higgs mass bound becomes 220 GeV
for $7 \lsim \tan\beta \lsim 25$ and reaches about 380 GeV for
$\tan\beta = 35$ for $\mu < 0$.
For $\mu > 0$ the \bsg process is not effective to put
constraints on the charged Higgs mass, just as in the case
of $m_t = 175$ GeV.

To summarize, we have shown that for $\mu<0$ the
lower bound of the charged Higgs mass increases 
compared to that in the Model \Rn{2} THDM. On the other
hand, for  $\mu>0$, the \bsg process cannot put useful
constraints on the allowed range of the charged Higgs mass
because it is possible that the charged Higgs
contribution is completely cancelled by other SUSY
contributions.
We have also pointed out that a parameter region corresponding to
the charged Higgs mass less than 180 GeV and 
$3 \lsim \tan\beta \lsim 5$ is excluded by the
\bsg process irrespective of the sign of $\mu$.

The authors would like to thank K.~Hikasa and T.~Yanagida
for reading the manuscript carefully and giving useful comments.
The work of Y.~O. is supported in part by the Grant-in-aid
for Scientific Research from the Ministry of Education,
Science and Culture of Japan.

%%%%%%%%%%%%%%%%%%%%%%%%%%%%%%%%%%%%%%%%%%%%%%%%%%%%%%%%%%%%%%%%%%%%%%

%%%%%%%%%%%%%%%%%%%%%%%%%%%%%%%%%%%%%%%%%%%%%%%%%%%%%%%%%%%%%%%%%%%%%%

\newcommand{\brCaption}{%
\bsg branching ratio for $m_t = 175$ GeV and $\tan\beta =
5$.
Each dot corresponds to a sample point which satisfies
radiative breaking and phenomenological constraints (see
text).
Solid line represents the branching ratio calculated with 
the SM and charged Higgs contributions only (Model \Rn{2} THDM).
Dot-dashed line represents the SM value.
}
\newcommand{\minusCaption}{%
Excluded region in the $\tan\beta$ and $m_{H^\pm}$ space for
$\mu < 0$.
Each line represents the lower bound for the charged Higgs mass;
solid line: all constraints included;
dashed line: without \bsg constraint;
dot-dashed line: Model \Rn{2} THDM with \bsg constraint.
}
\newcommand{\plusCaption}{%
Same as Fig.\ \protect\ref{fig:minus} for $\mu > 0$.
}
\newcommand{\excludeCaption}{%
Excluded region in the $\tan\beta$ and $m_{H^\pm}$ space
irespective of the sign of $\mu$.
The meanings of the lines are the same as those in Fig.\
\protect\ref{fig:minus}.
}
\newcommand{\lepIIminusCaption}{%
Excluded region in the $\tan\beta$ and $m_{H^\pm}$ space for
$\mu < 0$ with LEP \Rn{2} constraint.
The meanings of the lines are the same as those in Fig.\
\protect\ref{fig:minus}.
}
\newcommand{\lepIIplusCaption}{%
Same as Fig.\ \protect\ref{fig:lepIIminus} for $\mu > 0$.
}
\newcommand{\rminusCaption}{%
Excluded region in the $\tan\beta$ and $m_{H^\pm}$ space for
$\mu < 0$ when the initial condition for $\Delta_2^2$ is taken to
be $4\cdot m_0^2$.
The meanings of the lines are the same as those in Fig.\
\protect\ref{fig:minus}.
}
\newcommand{\rplusCaption}{%
Same as Fig.\ \protect\ref{fig:rminus} for $\mu > 0$.
}

\newpage
\section*{Figure Captions}

\begin{list}{\bf FIG.~??}{\relax}
\Fig{fig:br}         \brCaption
\Fig{fig:minus}      \minusCaption
\Fig{fig:plus}       \plusCaption
\Fig{fig:exclude}    \excludeCaption
\Fig{fig:lepIIminus} \lepIIminusCaption
\Fig{fig:lepIIplus}  \lepIIplusCaption
\Fig{fig:rminus}     \rminusCaption
\Fig{fig:rplus}      \rplusCaption
\end{list}

\newpage
\section*{Figures}

\begin{figure}[hbtp]
  \begin{center}
    \leavevmode
    \makebox[0cm]{
    \epsfile{file=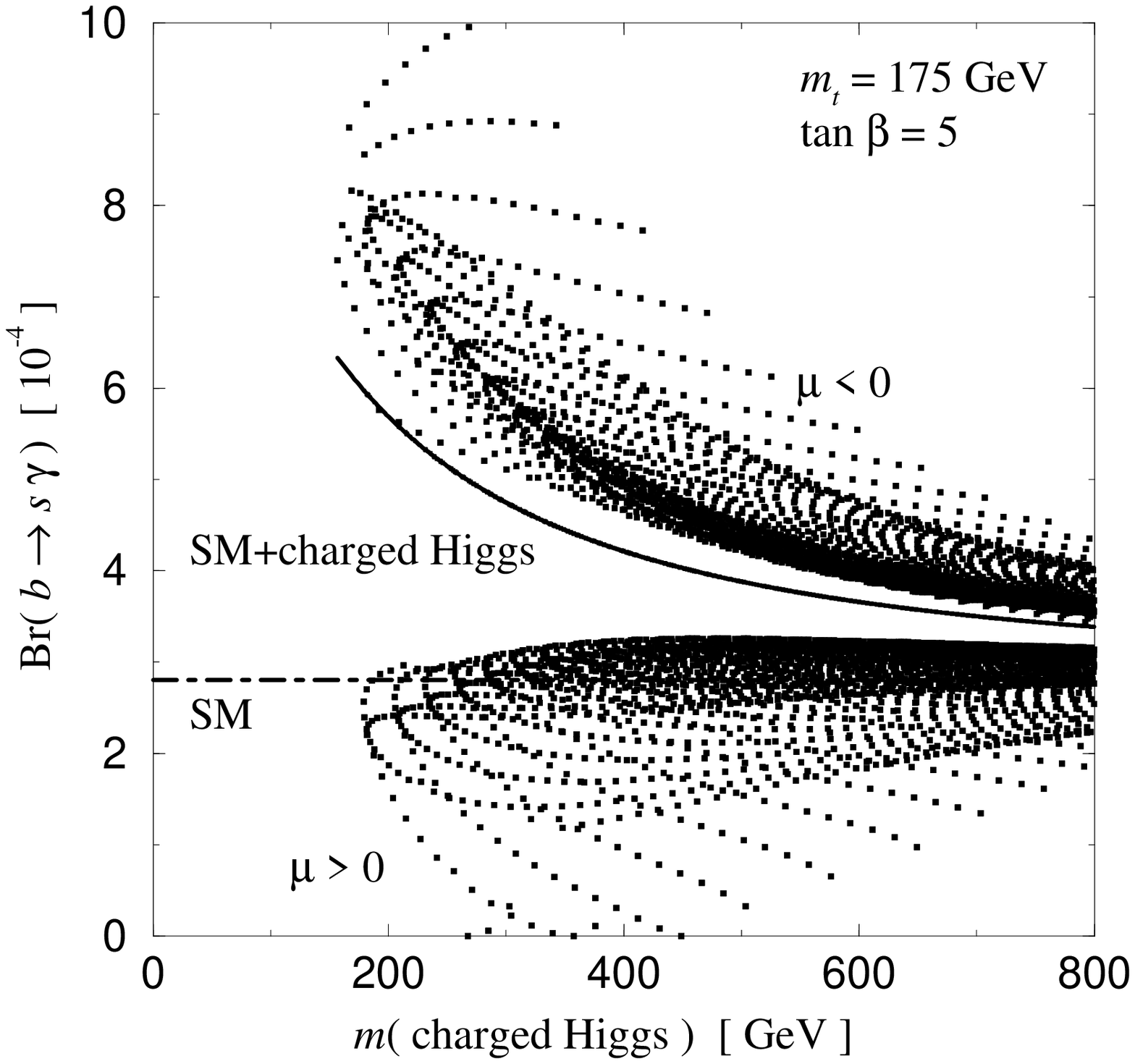,width=16cm}
    }
  \end{center}
  \caption{\brCaption}
  \label{fig:br}
\end{figure}

\begin{subfigures}

  \begin{figure}[p]
    \begin{center}
      \leavevmode
      \makebox[0cm]{
        \epsfile{file=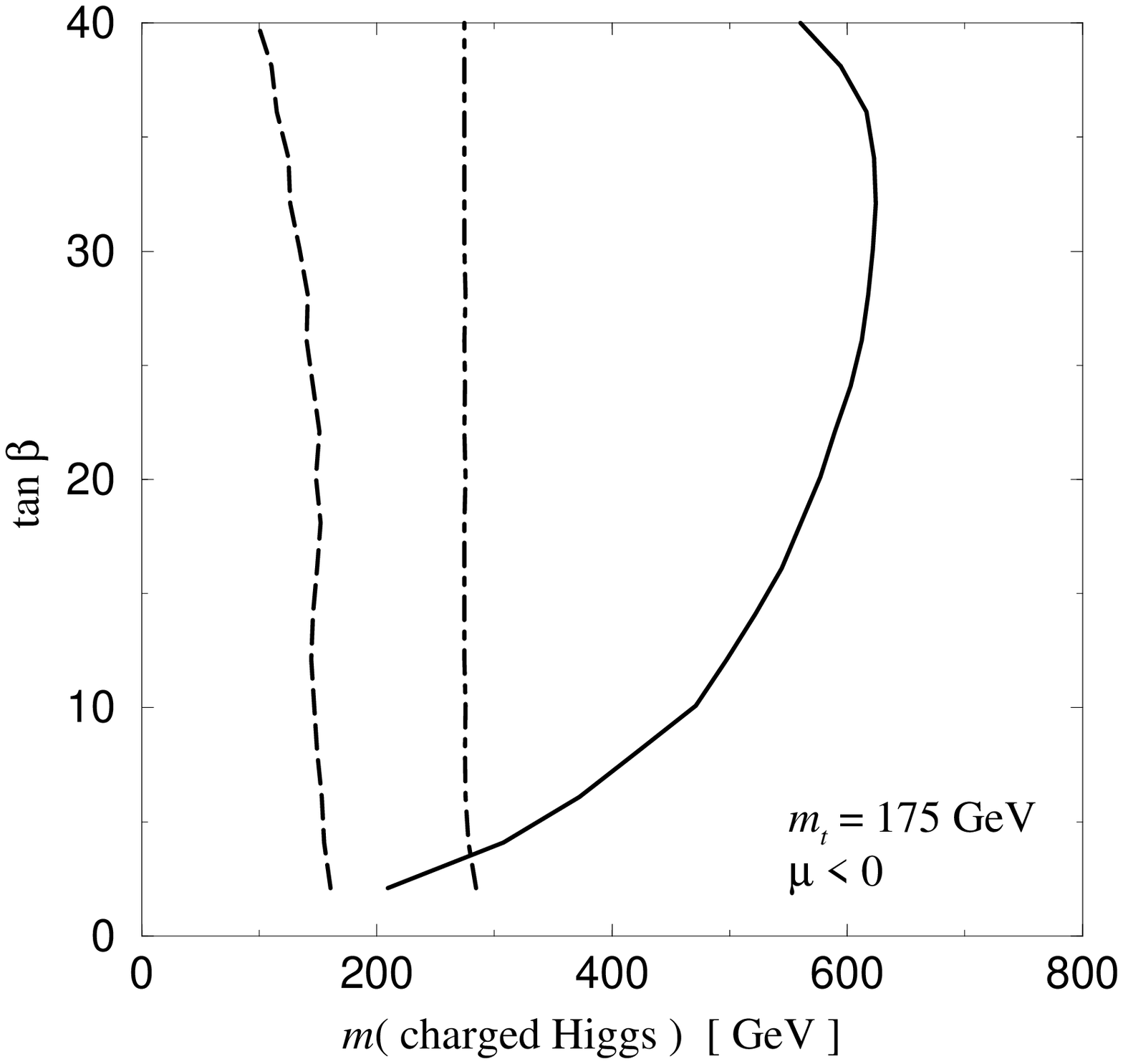,width=16cm}
        }
    \end{center}
    \caption{\minusCaption}
    \label{fig:minus}
  \end{figure}

  \begin{figure}[p]
    \begin{center}
      \leavevmode
      \makebox[0cm]{
        \epsfile{file=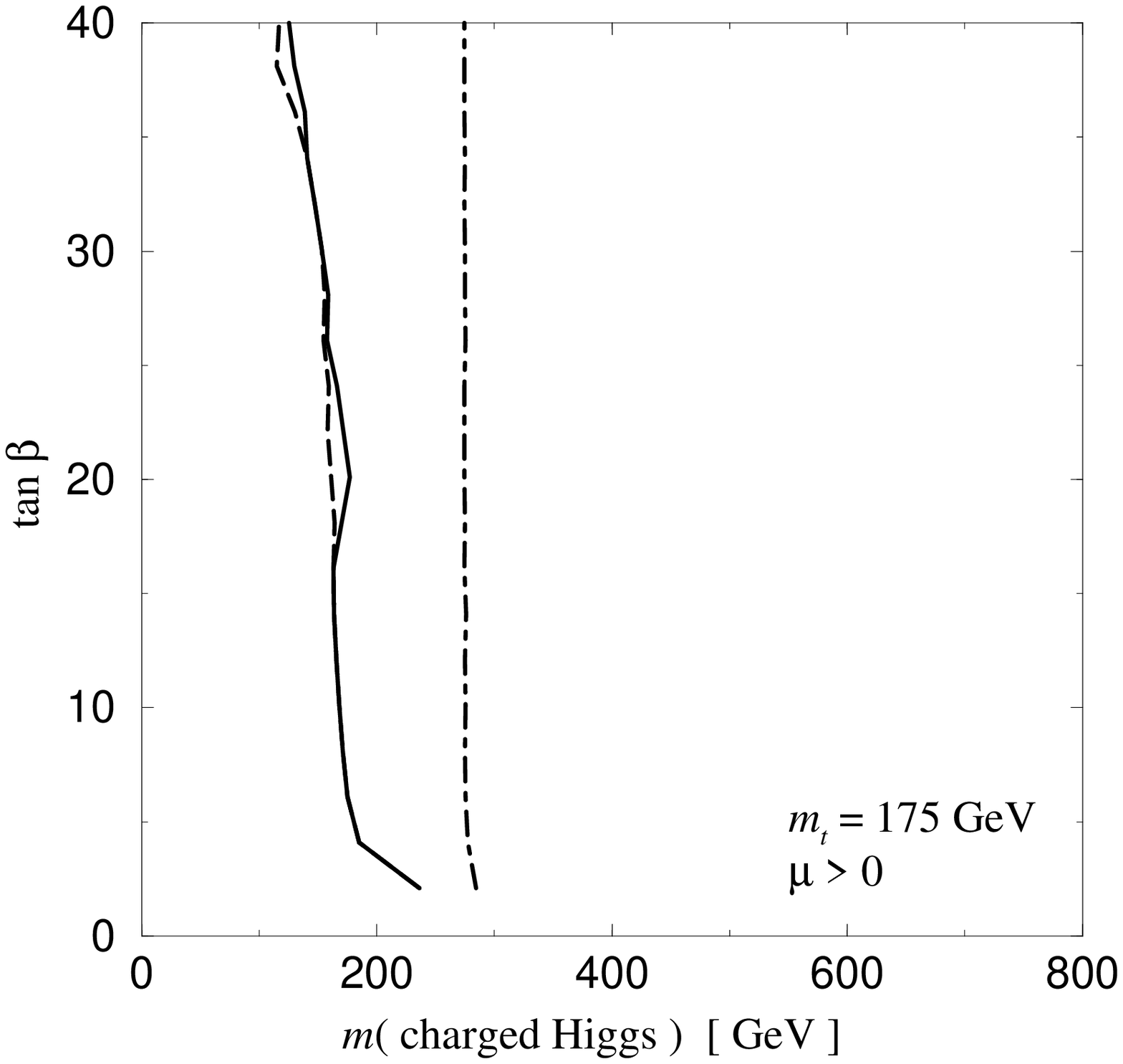,width=16cm}
        }
    \end{center}
    \caption{\plusCaption}
    \label{fig:plus}
  \end{figure}

\end{subfigures}

\begin{figure}[p]
  \begin{center}
    \leavevmode
    \makebox[0cm]{
      \epsfile{file=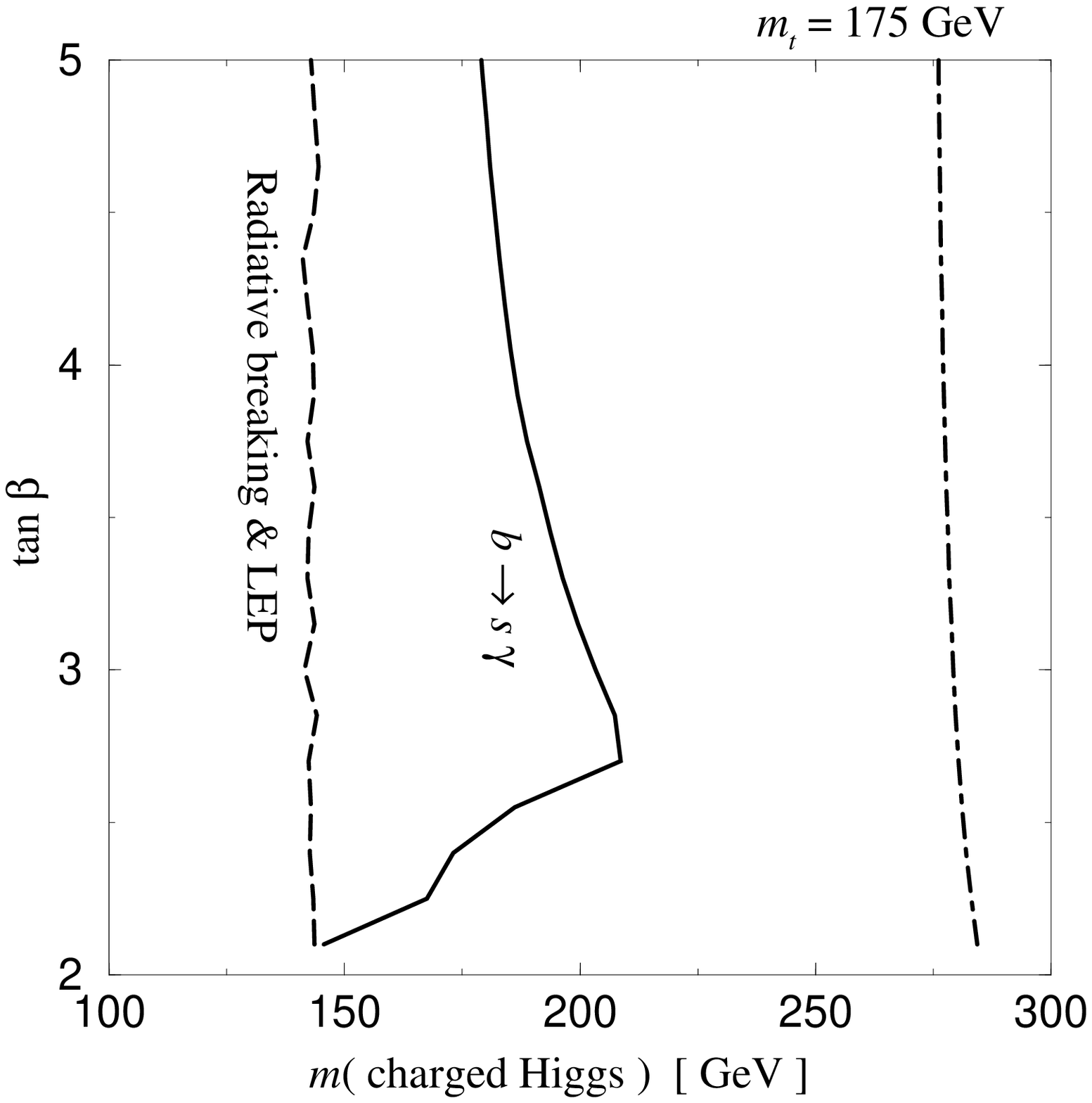,width=16cm}
      }
  \end{center}
  \caption{\excludeCaption}
  \label{fig:exclude}
\end{figure}

\begin{subfigures}

  \begin{figure}[p]
    \begin{center}
      \leavevmode
      \makebox[0cm]{
        \epsfile{file=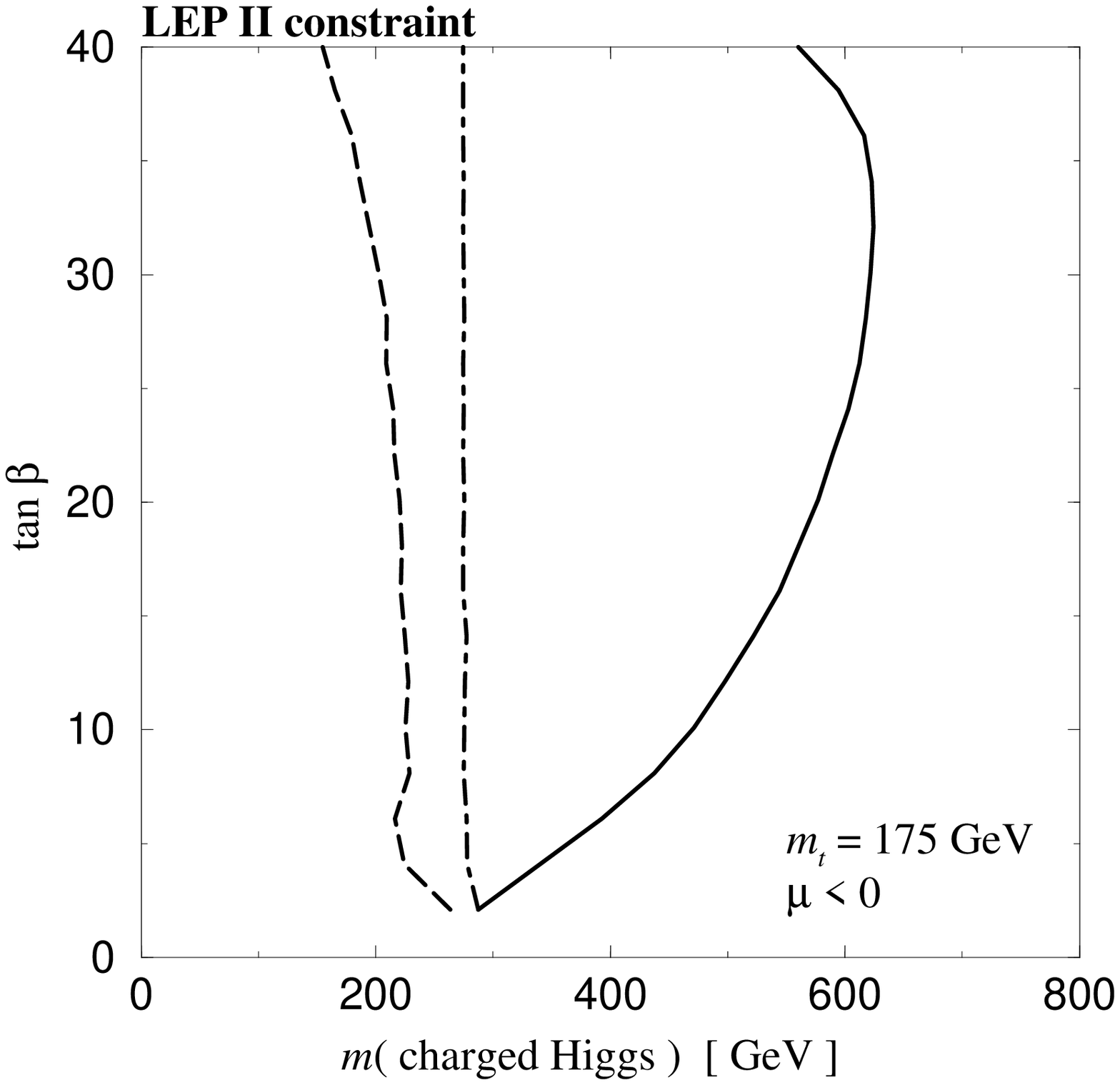,width=16cm}
        }
    \end{center}
    \caption{\lepIIminusCaption}
    \label{fig:lepIIminus}
  \end{figure}

  \begin{figure}[p]
    \begin{center}
      \leavevmode
      \makebox[0cm]{
      \epsfile{file=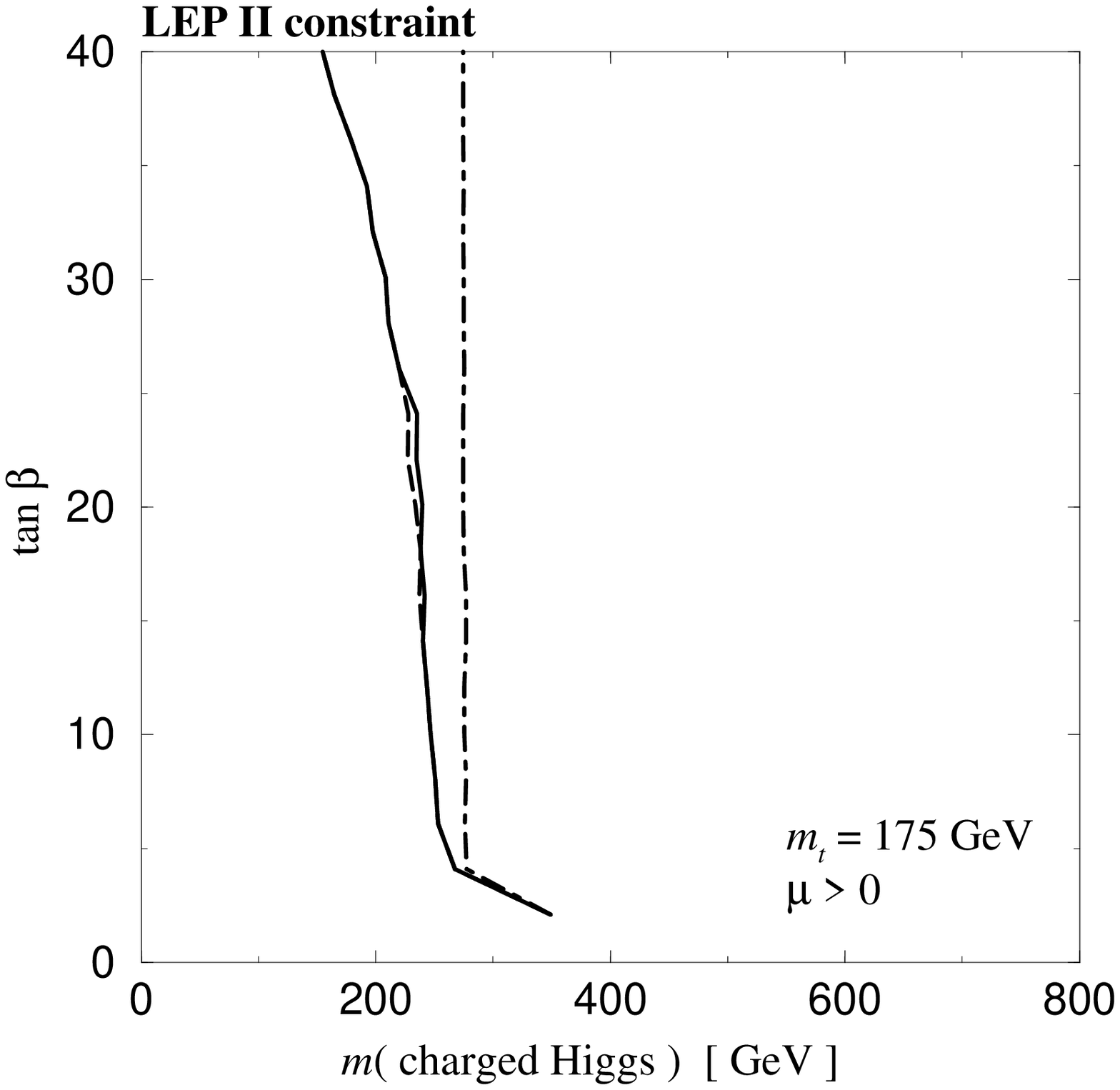,width=16cm}
      }
    \end{center}
    \caption{\lepIIplusCaption}
    \label{fig:lepIIplus}
  \end{figure}

\end{subfigures}

\begin{subfigures}

  \begin{figure}[p]
    \begin{center}
      \leavevmode
      \makebox[0cm]{
        \epsfile{file=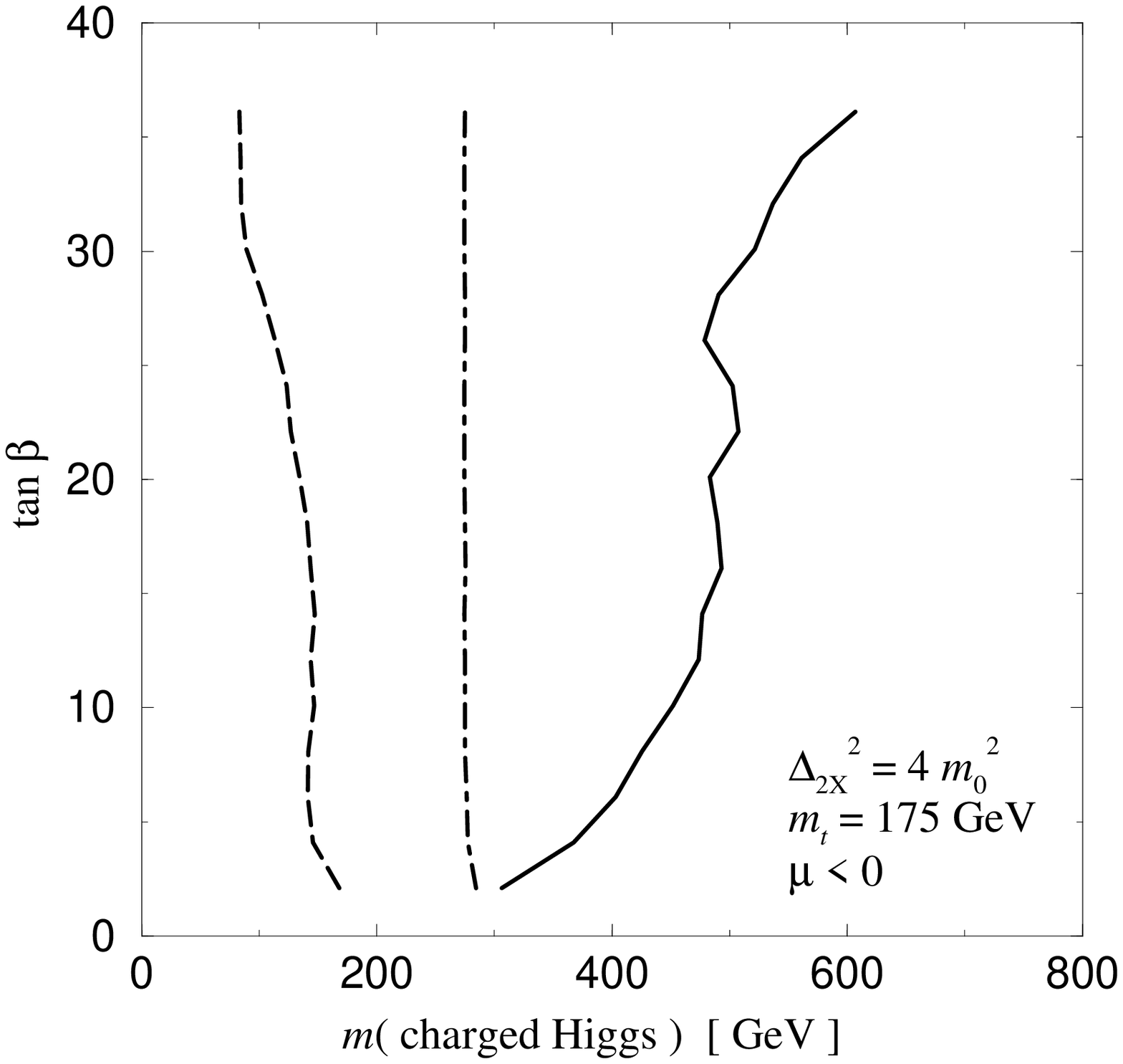,width=16cm}
        }
    \end{center}
    \caption{\rminusCaption}
    \label{fig:rminus}
  \end{figure}

  \begin{figure}[p]
    \begin{center}
      \leavevmode
      \makebox[0cm]{
        \epsfile{file=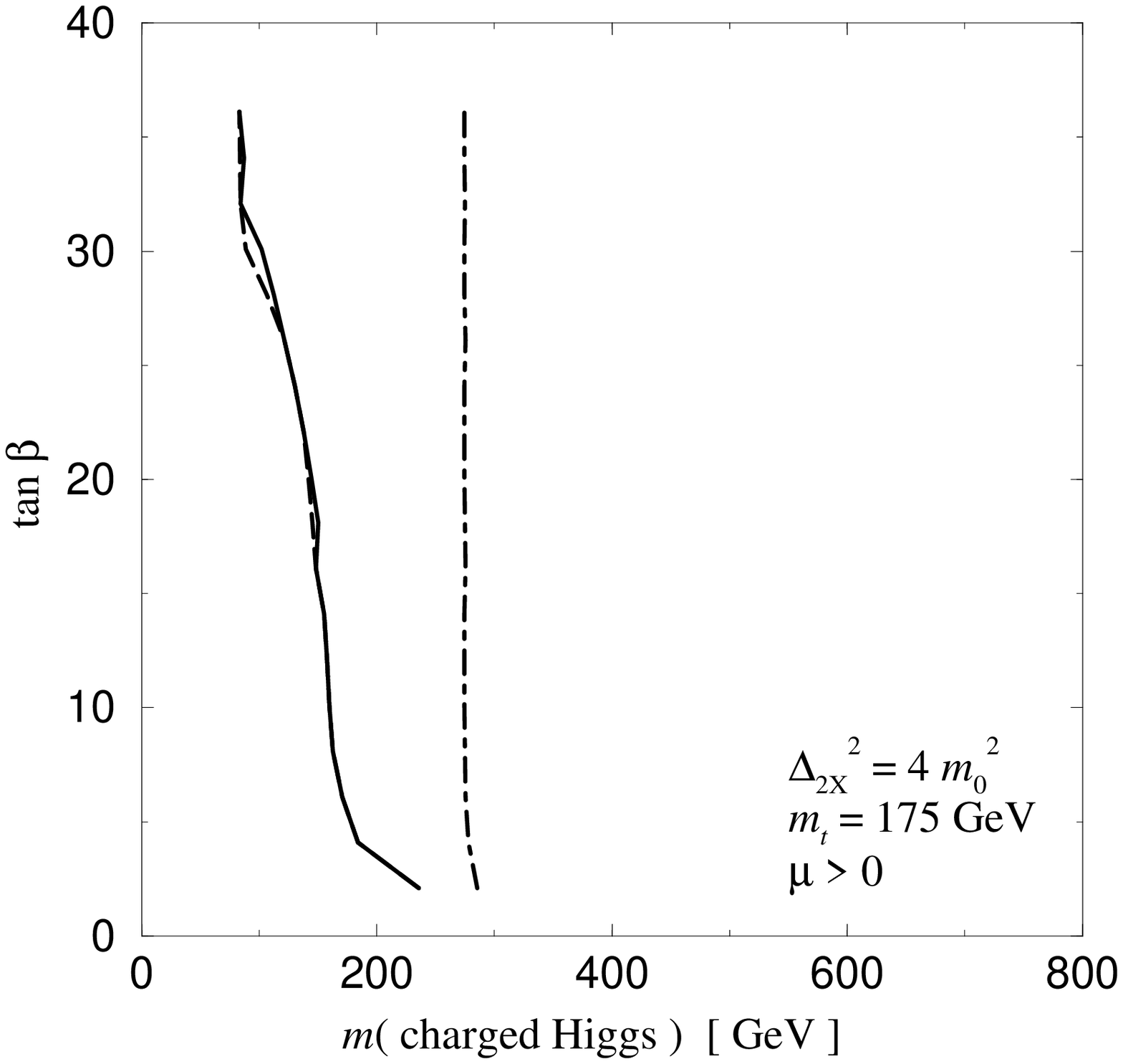,width=16cm}
        }
    \end{center}
    \caption{\rplusCaption}
    \label{fig:rplus}
  \end{figure}

\end{subfigures}

%%%%%%%%%%%%%%%%%%%%%%%%%%%%%%%%%%%%%%%%%%%%%%%%%%%%%%%%%%%%%%%%%%%%%%

\end{document}
%%%%%%%%%%%%%%%%%%%%%%%%%%%%%%%%%%%%%%%%%%%%%%%%%%%%%%%%%%%%%
%  EPS figures in tar-gzip-uuencoded form is the following. %
%%%%%%%%%%%%%%%%%%%%%%%%%%%%%%%%%%%%%%%%%%%%%%%%%%%%%%%%%%%%%
----------